\documentstyle[emulateapj, psfig]{article}

\lefthead{COORAY}
\righthead{UPPER LIMIT ON $\Omega_m$ USING LENSED ARCS}

\begin{document}
\title{An Upper Limit on $\Omega_m$ Using Lensed Arcs}
\author{Asantha R. Cooray$^1$}
\affil{$^1$Department of Astronomy and Astrophysics,
University of Chicago, Chicago IL 60637\\
E-mail: asante@hyde.uchicago.edu}


\begin{abstract}

We use current observations on the number statistics of
gravitationally lensed optical
arcs towards galaxy clusters to derive an upper limit on the
cosmological mass density of the Universe. 
The gravitational lensing statistics due to foreground clusters
combine properties of both cluster evolution, which is sensitive to
the matter density, and volume change, which is sensitive to 
the cosmological constant. The uncertainties associated  
with the predicted number of lensing events, however, currently do not
allow one to distinguish between flat and open cosmological models
with and without a cosmological constant. Still, after 
accounting for known errors,  and assuming that clusters in general have dark
matter core radii of the order $\sim$ 35 $h^{-1}$ kpc, 
we find  that the cosmological
mass density, $\Omega_m$, is less than 0.56 at
the 95\% confidence. Such a dark matter core radius is consistent with
cluster potentials determined recently by detailed numerical
inversions of strong and weak lensing imaging data. 
If no core radius is present, the upper limit on $\Omega_m$ 
increases to 0.62 (95\% confidence level).
The estimated upper limit on $\Omega_m$ is consistent with various 
cosmological probes that suggest a low matter density for
the Universe.

\end{abstract}

\keywords{cosmology: theory --- cosmology: observations ---
gravitational lensing}


\section{Introduction}

A large number of cosmological probes  now suggest that the Universe
is spatially flat with a low mass density (e.g., Perlmutter et al. 1998;
Riess et al. 1998; Lineweaver 1998; Guerra \& Daly 1998; Bahcall \& Fan
1998).  In addition to the mass density, gravitational lensing
statistics have allowed limits to be placed on the
cosmological constant. However, current limits on
the cosmological constant from gravitational lensing
arguments are only based on 
lensing statistics due to foreground galaxies (e.g., Kochanek 1996;
Falco, Kochanek, Munoz 1998; Cheng \& Krauss 1999; Cooray, Quashnock, Miller
1999; Cooray 1999a; Quast \& Helbig 1999)\footnote{We note that other
techniques, such as the luminosity distance to 
Type Ia supernovae at high redshifts (e.g., Perlmutter et al. 1998;
Riess et al. 1998), also allow constraints to be placed on the
cosmological constant}. 
An alternative approach is to consider
lensing statistics due to foreground galaxy clusters (e.g., 
Wu \& Hammer 1993; Bartelmann et al. 1998; Cooray 1999b).
It is well known that galaxy cluster evolution
is strongly sensitive to the cosmological mass density of the Universe (e.g.,
Bahcall \& Fan 1998; Viana \& Liddle 1998). Since lensing statistics
are sensitive to the cosmological constant, it is likely that
the number of lensed arcs due to galaxy clusters 
can provide strong constraints on both the mass density and the
cosmological constant.

Since the first suggestion that lensed optical arcs can be used as a 
cosmological probe (Wu \& Hammer 1993), several studies have addressed
specific issues related to the statistical calculation. These
include the effect of a cosmological constant (Wu \& Mao 1996) and 
background source evolution (Hamana \& Futamase 1997).
The numerical works by Bartelmann et al. (1998), using
simulated clusters in three cosmological models,  suggested that
current observational statistics on lensed
arcs are consistent with predictions in an open
Universe ($\Omega_\Lambda = 0$) with  $\Omega_m \sim 0.3$.
In Cooray (1999b; hereafter C99), we extended the predictions to
general cosmologies and also predicted the existence
of lensed radio and sub-mm towards foreground clusters.
Here, we extend the calculation in C99 by including various
uncertainties in the predicted number of lensed optical sources
to study the possibility of obtaining limits on cosmological
parameters based on the observed number.

In \S~2, we describe our calculation and inputs for the prediction.
In \S~3, we compare the predicted number of lensed arcs to the
observed number and use a reliable lower limit on the observed number to
derive an upper limit on the cosmological mass density of the
Universe. We follow the conventions that the Hubble constant,
$H_0$, is 100\,$h$\ km~s$^{-1}$~Mpc$^{-1}$, the present matter energy
density in units of the closure density is $\Omega_M$, and the normalized
cosmological constant is $\Omega_\Lambda$. Unless otherwise noted,
quoted errors are 1-$\sigma$ statistical errors.

\section{Gravitational Lensing Statistics}

In this section, we briefly describe our calculation and especially
the description of foreground lensing clusters (\S~2.1) and background
sources (\S~2.2). We also introduce a nonsingular isothermal
sphere model to describe galaxy cluster dark matter profile, which is primarily
motivated by recent determinations of the cluster potentials using high
performance numerical inversions of combined strong and weak lensing
data towards a sample of galaxy clusters. 

\subsection{Foreground Lenses}

The differential probability that a beam towards a background source 
will encounter a foreground lens with a path length of $dz_L$ is:
\begin{equation}
d\tau = n(z_L) a_{\rm lens} \frac{c dt}{dz_L}dz_L,
\end{equation}
where $n(z_L)$ is the number density of foreground lenses at redshift
$z_l$ while $a_{\rm lens}$ is the lensing cross section (e.g.,
Fukugita et al. 1992). 

Using the Press-Schechter mass function (Press \& Schechter 1974; PS),
the comoving number density of galaxy clusters, $dn(M,z)$, at
redshift $z$ and mass $(M,M+dM)$, can be written as
\begin{eqnarray}
\frac{dn(M,z)}{dM} = \\ \nonumber
- \sqrt{\frac{2}{\pi}} \frac{\bar{\rho}}{M}\frac{d\sigma(M,z)}{dM}
\frac{\delta_{c}}{\sigma^2(M,z)}
\exp{\left[\frac{-\delta_{c}^2}{2 \sigma^2(M,z)}\right]}\, , \nonumber
\end{eqnarray}
where $\bar{\rho}$ is the comoving background matter density,
$\sigma^2(M,z)$ is the variance of the fluctuation spectrum averaged over
a mass scale $M$, and  $\delta_{c}$  is the
linear overdensity of a perturbation which has collapsed and virialized.
Taking an approach similar to the one presented in Viana \& Liddle
(1998), $\sigma(M,z)$ is written as a function of the comoving radius,
$R$, which contains mass $M$ at the current epoch:
\begin{equation}
\sigma(R,z)=\sigma _8(z) \left({ R \over 8 h^{-1} 
           {\rm Mpc}}\right)^{-\gamma(R)},
\end{equation}
where
\begin{equation}
\gamma(R) = (0.3\Gamma + 0.2)\left[2.92 + \log_{10}
\left({R \over 8 h^{-1} {\rm Mpc}}\right)\right] \, .
\end{equation}
Here, $\Gamma=0.23 \pm 0.05$ (Peacock \& Dodds 1994) 
is the CDM shape parameter; our results are insensitive to its
specific value (e.g., Viana \& Liddle 1998).  
In order to calculate growth evolution as a function of
redshift  in various cosmologies, we write $\sigma_8(z)$ as:
\begin{equation}
\sigma_8(z) = {\sigma_8(0)\over 1+z}\,\,
{g(\Omega_m(z)) \over g(\Omega_m(0))} \, ,
\end{equation}
where, following Carroll, Press \& Turner
(1992), the growth suppression factor is:
\begin{equation}
g(\Omega_m) = {5 \over 2} \Omega_m
\left[ \Omega_m^{4/7} - \Omega_{\Lambda} +
\left(1 + {\Omega_m \over 2} \right)
\left(1 + {\Omega_{\Lambda} \over 70} \right) \right]^{-1} \, .
\end{equation}

The normalization for $\sigma_8$ comes from the local temperature
function (Pen 1998):
\begin{equation}
\sigma_8(0) = \left\{ \begin{array}{cl}
  (0.53 \pm 0.05) \, \Omega_m^{-0.46} & {\rm \Omega_\Lambda=0\,} \\
  (0.53 \pm 0.05) \,\Omega_m^{-0.53} & {\rm \Omega_m+\Omega_\Lambda=1\, .} \\
 \end{array} \right.
\end{equation}

In order to model the cluster lensing potential, we use the
nonsingular singular
isothermal sphere model with the observed velocity
dispersion and an a priori determined value for the core radius of the
dark matter potential of the cluster. The evidence for
a core radius in the dark matter profile of galaxy clusters primarily 
comes from the existence of gravitationally lensed arcs in the radial direction
from the cluster center. For simple models for the cluster potential
involving
singular isothermal models, such arcs are located at a distance
equivalent to the core radius of the cluster pontential profile.
Also, recent numerical inversions of galaxy cluster lensing potentials
using Hubble Space Telescope and other ground based high quality images
clearly suggest the presence of a small core
radius (Tyson, Kochanski, Dell'Antonio 1998; Ian Dell'Antonio, 
private communication). Thus, it is necessary that we consider a lensing model 
which allows for the possible presence of a core radius. Following Hinshaw \&
Krauss (1987), we consider a isothermal sphere model with a core
radius and write the density profile as:
\begin{equation}
\rho = \frac{\sigma_{\rm vel}^2}{2 \pi G(r^2+r_c^2)},
\end{equation}
where $\sigma_{\rm vel}$ is the dark matter velocity dispersion
and $r_c$ is the core radius of the dark matter profile of the
cluster. The conventional singular isothermal sphere (SIS) is
recovered when $r_c$ is zero. 
The lensing cross section for the nonsingular isothermal model 
is given by:
\begin{equation}
a_{\rm lens} = 16 \pi^3 \left( \frac{\sigma_{\rm vel}}{c} \right)^4
\left( \frac{D_{OL}D_{LS}}{D_{OS}}\right)^2 f(\beta)
\end{equation}
where $D_{OL}$, $D_{OS}$ and $D_{LS}$ are observer to lens,
observer to source and lens to source distances. These distances are
calculated under the filled beam approximation. 
In Eq.~9, $f(\beta)$ is a correction factor that takes into account
the nonsingular behavior of the density profile (see, Hinshaw \&
Krauss 1987):
\begin{equation}
f(\beta) = 1 +5\beta-\frac{\beta^2}{2} -
\frac{\sqrt{\beta}(4+\beta)^{3/2}}{2},
\end{equation}
where $\beta$ is the ratio of core radius to critical radius of the
lensing potential, with the latter measured at the redshift of the
cluster:
\begin{equation}
\beta  = \frac{r_c c H_0 (1+z_L)}{4 \pi \sigma_{\rm vel}^2} \left(
\frac{D_{OS}}{D_{LS}D_{OS}}\right).
\end{equation}
When the SIS model is considered, $\beta = 0$ and $f(\beta)=1$.
For small core radii, especially for the present case involving
galaxy clusters, one can usually ignore higher order $\beta$ 
terms associated with $f(\beta)$; we consider, however, the full
formula in deriving cosmological parameters.
Finally, the differential optical depth for the nonsingular isothermal
model is: 
\begin{eqnarray}
d\tau = 16 \pi^3 \left(\int_{M_{\rm min}}^{\infty}
\left[\frac{\sigma_{\rm vel}(M')}{c}\right]^4 \frac{dn(M',z_L)}{dM'}\,
f(\beta) dM'\right)  \\ \nonumber
\times
 (1+z_L)^3 \left( \frac{D_{OL}D_{LS}}{D_{OS}}\right)^2 \frac{cdt}{dz_L}dz_L,
\end{eqnarray} 
The total
optical depth to a given background redshift, $z_s$, is given by:
\begin{equation}
\tau(z_s) = \int_{0}^{z_s} \frac{d\tau}{dz_L}dz_L\;.
\end{equation}

In order to calculate the lensing optical depth, we take a two step
approach to relate cluster velocity dispersion to its mass.
We relate velocity dispersion to cluster temperature
using  recently updated $\sigma-T$ relation (Wu et al. 1998):
\begin{equation}
\sigma_{\rm vel}(T) = 10^{2.57 \pm 0.03} \left(\frac{T}{{\rm keV}}\right)^{0.56
\pm 0.09}\; {\rm km\; s^{-1}};
\end{equation}
and then to mass  using partly theoretical 
$M-T$ relation (e.g., Barbosa et al. 1996):
\begin{eqnarray}
T(M,z) = (6.8 \pm 0.5) h^{\frac{2}{3}}\; {\rm keV}\; 
\left[\frac{\Omega_m \Delta_c(\Omega_m,z)}{178}\right]^{\frac{1}{3}} 
\\ \nonumber
 \times \left(\frac{M}{10^{15}h^{-1} M_{\sun}}\right)^{\frac{2}{3}}(1+z).
\end{eqnarray}
We have allowed for an extra 
uncertainty in the $M-T$ relation by comparing various
 normalizations that have been suggested in the literature.
Also, we note that $\beta$ is dependent on cluster mass through 
velocity dispersion (Eq.~11). In addition to velocity dispersion, it is likely
that cluster core radii are also dependent on individual cluster
masses. Even though such variations have been 
observationally determined for galaxies
(e.g., Lauer 1985), there is still no observational evidence
for a dependence of galaxy cluster core radii with other physical properties,
such as the X-ray luminosity or temperature. For the purpose of this
calculation, we take a constant value for the core radius based on the
mean value of cluster core radii from numerical 
inversions ($\sim$ 35 h$^{-1}$ kpc; Dell'Antonio private
communication).  
When deriving cosmological parameters, we vary the exact value 
of the core radius to investigate the parameter dependences 
on it; As we find later, our limits on
$\Omega_m$ is weakly dependent on the core radius. 

\vskip 2mm 
\hbox{~}

\centerline{\psfig{file=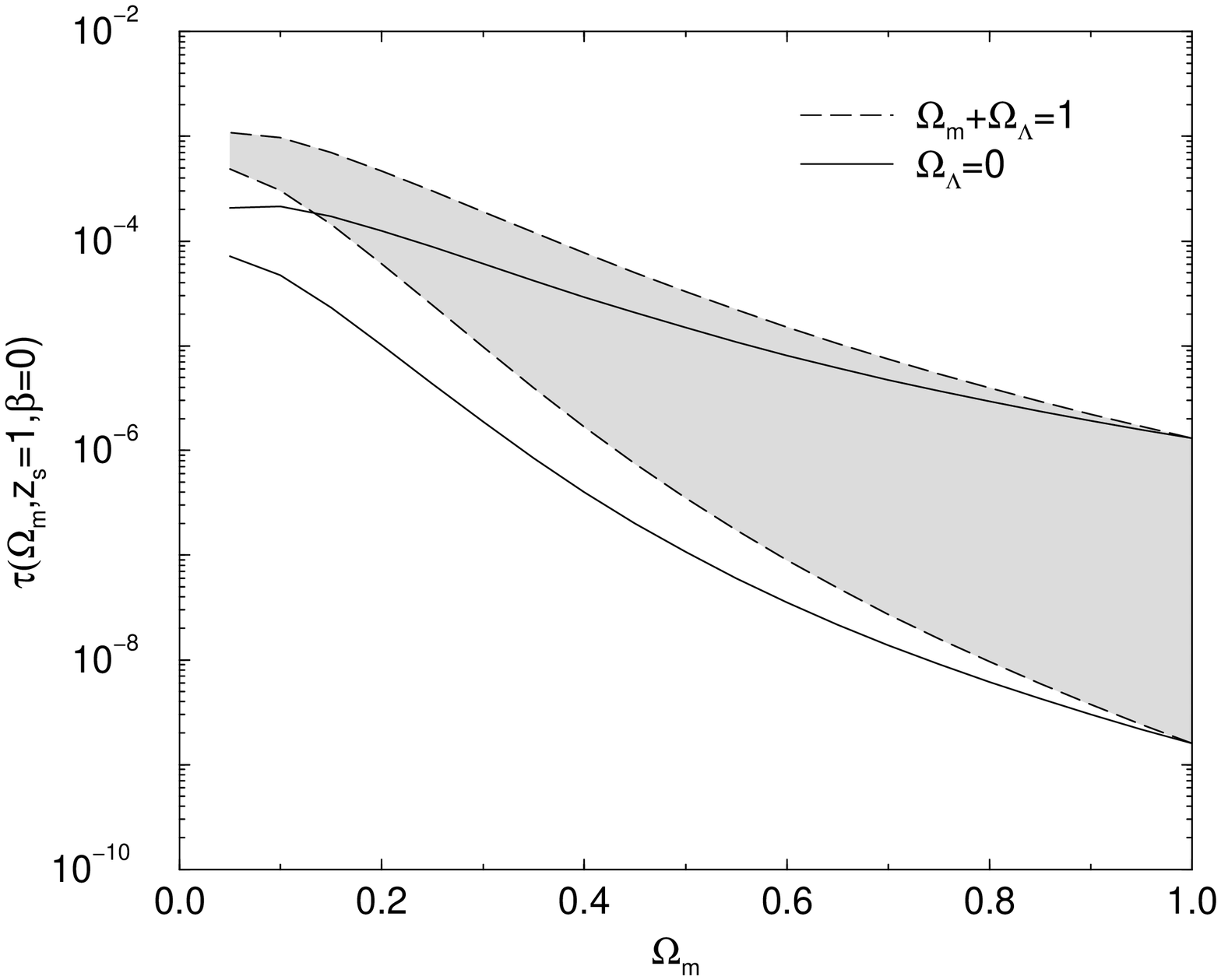,width=4.0in,angle=-90}}
\noindent{
\addtolength{\baselineskip}{-3pt} 
\vskip 1mm

Fig.~1.\ Optical depth for strong lensing for a background source
at redshift of 1, due to foreground massive clusters. Shown are
the 95\% confidence ranges for both flat ({\it dot-dashed lines}) and open 
({\it solid lines}) cosmologies
with and without a cosmological constant. Given the large uncertainty
associated with the optical depth and the small difference between
flat and open cosmological models, 
it is unlikely that lensed arc statistics can be used to reliably
place limits on the cosmological constant.

\vskip 3mm
\addtolength{\baselineskip}{3pt}
}

In Fig.~1, as an illustration, 
we show the optical depth to lensing due to foreground
clusters with total masses greater than 
$7.5 \times 10^{14}\; h^{-1}\; {\rm M_{\sun}}$ for a 
background source at a redshift of 1, as a function of $\Omega_m$ for
open and flat cosmologies, and considering a lensing potential
in which $f(\beta)=1$ (SIS model). Shown are the 95\% upper and lower
confidences in each case by considering all possible errors we
have so far considered.  The uncertainty in the optical depth is 
primarily dominated by the error associated with the normalization of 
the PS mass function;  since the number density of massive clusters is strongly
sensitive to the exponential term in Eq.~2,
small changes in $\sigma_8$ can produce order of magnitude
changes in the number density. The difference between flat and open
cosmological models is primarily due to the increase in lensing probability
with the addition of $\Omega_\Lambda$. However, this difference
is small, and when errors in observations are also
considered, it is impossible to study the possible
existence of a cosmological constant
using lensed arc statistics. Therefore, 
taking a conservative approach, we combine the upper curve valid for flat
cosmologies with the lower curve defined by open models to
combine the 95\% confidence range in the predicted number of lensed sources.

\subsection{Background Sources}

In order to obtain reliable predictions on the number of lensed arcs,
it is important that both the background source evolution
and effects such as ``magnification bias'' (Kochanek 1991) be
included in the calculation. Our description of background sources
comes from the Hubble Deep Field (HDF; Williams et al. 1996). We use the
HDF redshift and magnitude 
distribution and the luminosity function from Sawicki et al. (1997).
Such an approach allows us to reliably account for the true 
redshift distribution of 
background sources, instead of an empirical distribution
or a constant redshift, while also accounting for 
intrinsic evolutionary effects which has shown to be
important for lensing predictions (e.g., Hamana \& Futamase 1997) 

Using the probability,
$\tau(z,\Omega_m,M_{\rm min})$,
 for a source at redshift $z$ to be strongly lensed 
and the number  of unlensed background sources  between rest-frame
luminosity $L$ and  $L+dL$ and between redshifts $z$ and $z+dz$,
$\Phi(L,z)dL\,dz$, we can write the number of lensed galaxies,
 $d\bar N$, in that luminosity and 
redshift interval as (see, also Maoz et al.\ 1992):
\begin{eqnarray}
{d\bar N(L,z)\over dz}=\tau(z,\Omega_m,M_{\rm min}) \\ \nonumber
\times \int\left[\Phi\left({L\over A},z\right)\,
{dL\over A}\right]f(A,L,z)q(A)\,dA \;.
\end{eqnarray}
Here, the integral is over all allowed values of $A$, the amplification
of the brightest lensed image,
$q(A)$ is the probability distribution of amplifications, and
$f(A,L,z)$ is the probability of observing the brightest image
given $A$, $L$, and $z$.  Our assumption that the lenses are nonsingular
isothermal spheres implies that the minimum amplification, $A_{\rm
min}$, is a function of $\beta$. In general, the probability
distribution of amplifications can be written as:
\begin{equation}
q(A)\,dA=2 A_{\rm min}^2 A^{-3}\,dA.
\end{equation}

\vskip 2mm 
\hbox{~}

\centerline{\psfig{file=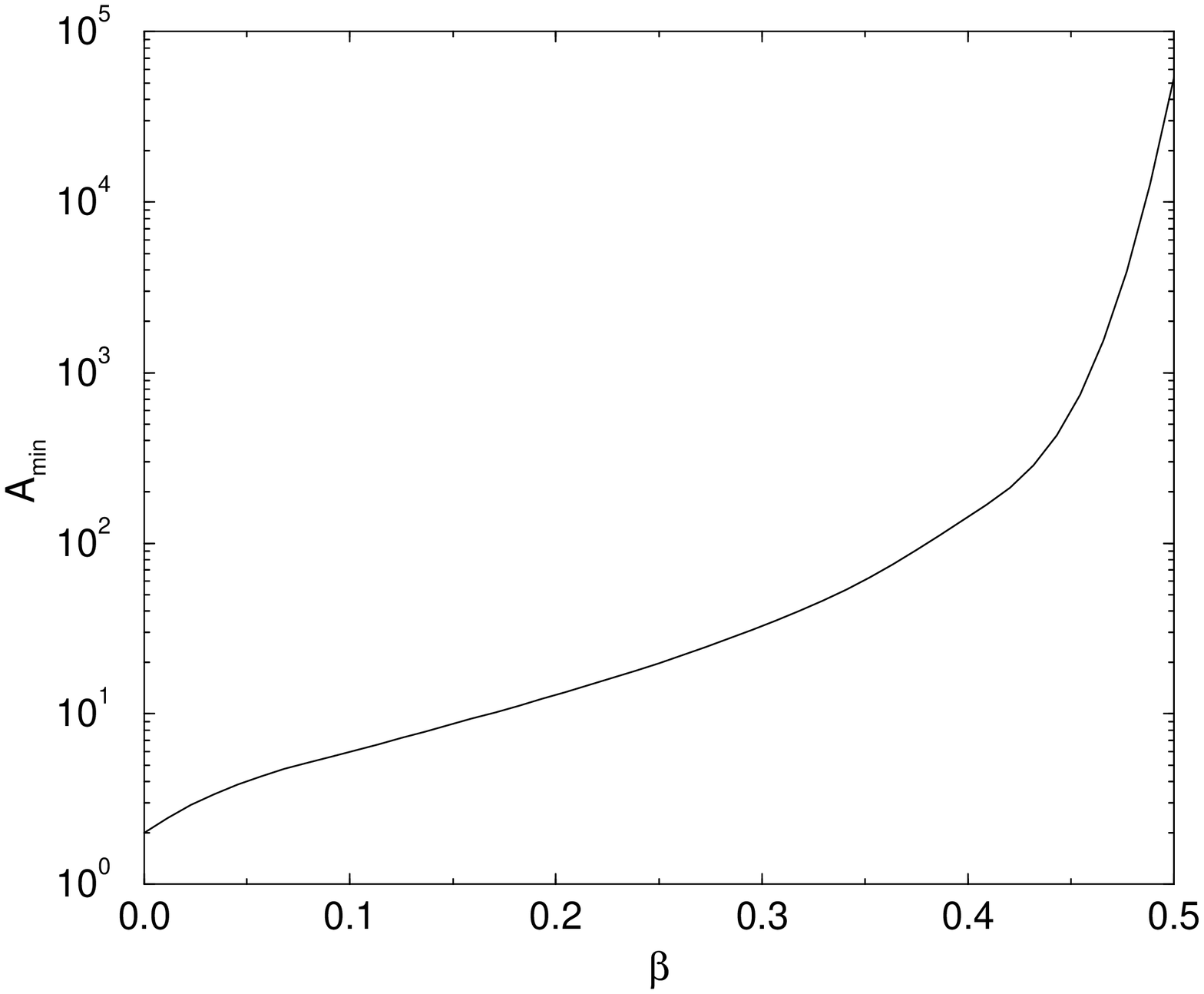,width=4.0in,angle=-90}}
\noindent{
\addtolength{\baselineskip}{-3pt} 
\vskip 1mm

Fig.~2.\ Minimum amplification versus 
$\beta$, the ratio of core radius to
critical radius at the redshift of the lensing cluster.

\vskip 3mm
\addtolength{\baselineskip}{3pt}
}

In Fig.~2, we show $A_{\rm min}$ as a function of $\beta$, which is
calculated following Cheng \& Krauss (1999).
In practice we use a fitting function that returns $A_{\rm min}$
for a given value of $\beta$, with an accuracy of better than 0.1\%
at all interested values of $\beta$ in the present calculation.
For simplicity, we assume that 
$f(A,L,z)$ is a step function, $\Theta[m_{\rm lim},A]$, so that a
lensed image with apparent magnitude brighter than $m_{\rm lim}$
is detected. For a given value of the core radius and the
velocity dispersion, $\beta$ is determined from
Eq.~11. For massive clusters
discussed here with velocity dispersions of the order $\gtrsim$ 
1000 km s$^{-1}$ and at redshifts $\sim$ 0.2, $\beta \lesssim 0.07$
and $A_{\rm min} \lesssim 5$. Compared to
the SIS model, the addition of a small core radius only produces 
slight changes in the lensing probability. As
described in Kochanek (1995), the effect of a core radius
is to increase the magnification bias while increasing the effective 
lensing cross section; the overall effect is that the 
presence of a core radius is not significantly different from that 
of a SIS model.

We assume that the brightness distribution of background galaxies
at any given redshift is described by a Schechter function (Schechter
1976),  in which the comoving density of galaxies at redshift $z$ and with
luminosity between $L$ and $L+dL$ is
\begin{equation}
\phi(L,z)\, dL=\phi^*(z)\left[L\over{L^*(z)}\right]^{\alpha(z)}
e^{-L/L^*(z)}\, dL\; ,
\end{equation}
where, as before, both $L$ and $L^*$ are measured in the rest
frame of the galaxy.  Following CQM, 
we can  write the expected number $\bar N$ of lensed sources as
\begin{eqnarray}
\bar N &=\sum_i \tau(z_i,\Omega_m,M_{\rm min})
 \int_2^\infty A^{-1-\alpha(z_i)}e^{L_i/L^*(z_i)}
e^{-L_i/AL^*(z_i)} \nonumber \\
&\times \Theta\left[m_{\rm lim},A\right]
{2\over{(A-1)^3}}dA\;,
\end{eqnarray}
where the sum is over each of the background galaxies.
The index $i$ represents
each galaxy; hence, $z_i$, $L_i$, and $m_i$ are, respectively,
the redshift, rest-frame luminosity, and apparent magnitude
of the $i$th galaxy.

Since $L_i$ for individual galaxy is unknown, due to uncertain
K-corrections, following Cooray, Quashnock \& Miller (1999),
we estimate the total average bias 
by weighting the integral in Eq.~19 by a normalized
distribution of luminosities $L_i$ drawn from the Schechter function
appropriate for the redshift $z_i$ of galaxy $i$.
We calculated the magnification bias for 
individual redshift intervals for which the Schechter function
parameters are available in Table 1 of Sawicki et al.
(1997).  In principle, the uncertainties
in the Schechter function parameters at a given redshift can affect
the calculation of the bias, but in practice only the
uncertainty in the power-law slope $\alpha$ has a significant effect.
The effect of varying $\alpha$ on the lensing statistics was discussed
in Cooray, Quashnock \& Miller (1999) for lensing statistics involving
foreground galaxies in the Hubble Deep Field, 
which is also valid for the present case
involving galaxy clusters; the general
effect due to uncertainties tabulated in Sawicki et al. (1997) is that the
constraints on cosmological parameters vary by less
than 5\% percent, when $\alpha$ is in general varied by the quoted
1-$\sigma$
uncertainties in Sawicki et al. (1997).
Here, we take a conservative approach
and allow the largest possible bias, so that the expected number is
overestimated by an amount as suggested above. The only effect of this
approach is to 
slightly increase our upper limit on $\Omega_m$. 

\vskip 2mm 
\hbox{~}

\centerline{\psfig{file=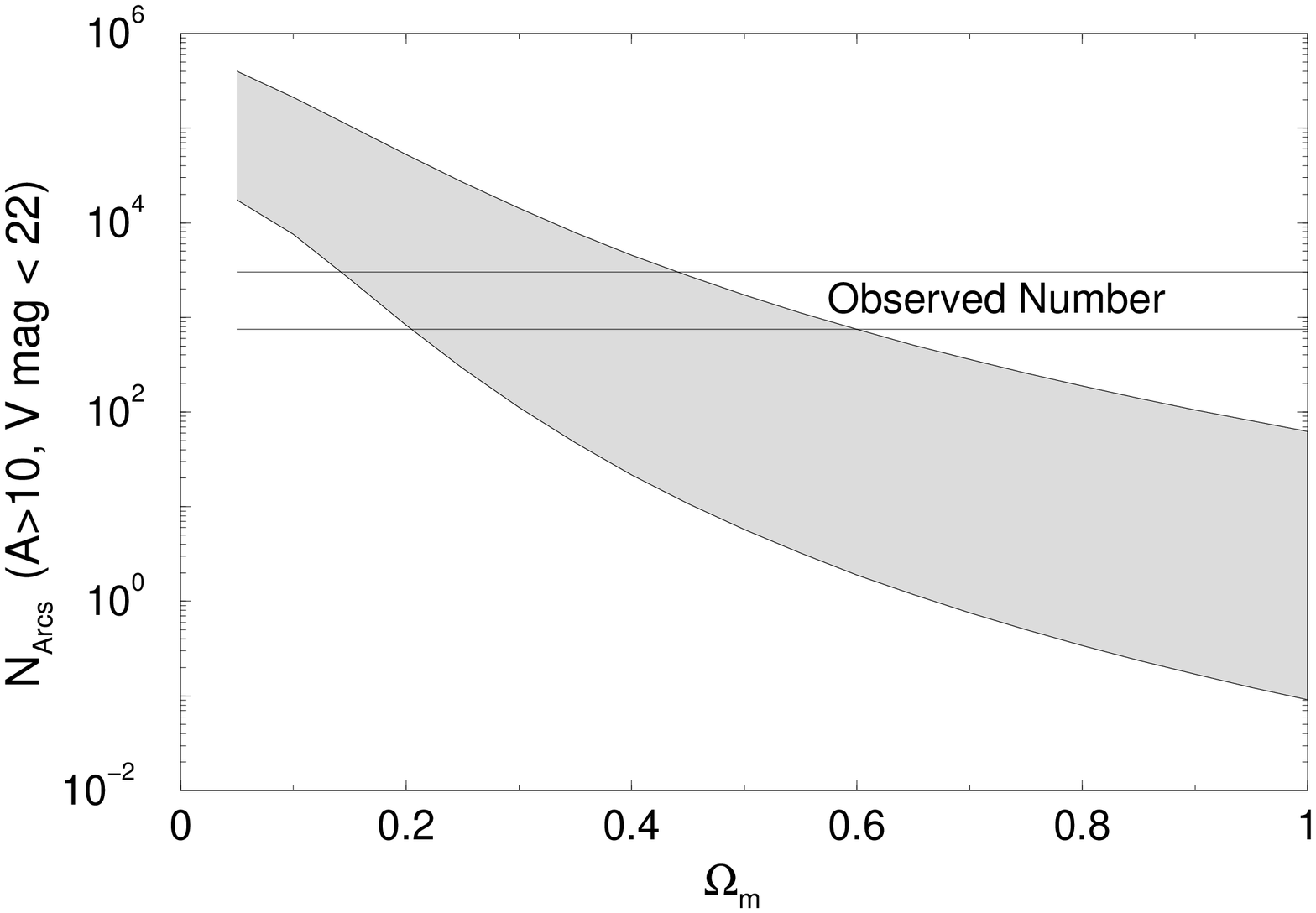,width=4.0in,angle=-90}}
\noindent{
\addtolength{\baselineskip}{-3pt} 
\vskip 1mm

Fig.~3.\ Expected number of lensed arcs on the whole sky with
amplifications
greater than 10 and V-band magnitudes brighter than 22 towards
foreground massive clusters. The shaded range shows the 95\%
confidence upper and
lower limits on the expected number of lensed arcs, while the
horizontal lines show the range of current observed numbers. We use the
lower limit on the current observed number to impose an
upper limit on $\Omega_m$.
\vskip 3mm
\addtolength{\baselineskip}{3pt}
}

In Figure~3, we show the expected number of lensed arcs towards
foreground massive clusters with total mass greater than $M_{\rm min}
= 7.5 \times 10^{14}\; h^{-1} {\rm M_{\sun}}$, and assuming a zero
core radius for the lensing potential.
We define an arc as a lensed source which is amplified by a factor
equal to or greater than 10. To make a direct
comparison to both observations and prior predictions, we impose
a limiting V-band magnitude of 22. Our numbers can be directly
compared to previous estimates, especially those of Bartelmann et al.
1998). This study predicted $\sim$ 2400 arcs in an open Universe with
$\Omega_m \sim 0.3$ and $\sim$ 36 arcs in an Einstein-de Sitter Universe.
The number expected in a flat Universe with $\Omega_m \sim 0.3$
and $\Omega_\Lambda \sim 0.7$ was $\sim$ 280. Our estimates for
for an Einstein-de Sitter Universe range from $\sim$ 0.1 to 60
while for $\Omega_m \sim 0.3$  Universe (independent of
$\Omega_\Lambda$) is $\sim$ 50 to 7000 (with the higher end allowed by
the cosmological constant). The primary reason for a lower number of
arcs with $\Omega_\Lambda$  in the study by Bartelmann et al. (1998) 
was their assumption that clusters are different in Universes with
 a cosmological constant, such that their concentration is lower.
Based on numerical simulations performed by the Virgo Consortium, however,
Thomas et al. (1998) studied a series of clusters in four different
cosmologies, including an open model with $\Omega_m=0.3$ and
a flat model with a cosmological constant of $\Omega_\Lambda=0.7$.
The authors concluded that clusters do not exhibit differences
between open and flat cosmologies with and without
a cosmological constant and that cluster structures
cannot be used to discriminate between the two possibilities.
If Thomas et al. (1998) are correct, then
the inclusion of a cosmological constant is not expected to change cluster mass
profiles to an extent that would affect the gravitational lensing rate.
In any case, such systematic effects are unlikely to
be nearly as large as the current uncertainty in $\sigma_8$ which
dominates the present calculation on the lensing rate.
Ignoring this case, our predictions are
generally consistent with Bartelmann et al. (1998). 
As we have demonstrated in Fig.~1, lensed arc statistics are unlikely
to provide useful limits on the cosmological constant.
The same is true for alternatives
to the cosmological constant, such as scalar field and quintessence
models  that have recently been introduced 
(e.g., Steinhardt et al. 1998).

\section{Constraints on $\Omega_m$}

In order to derive a limit on $\Omega_m$ based on the number of lensed
arcs, we require knowledge on the observed number of such lensing
events. Current surveys of clusters are based on their X-ray
luminosities rather than masses. For an example, the luminosity
cutoff of the followup Einstein Medium Sensitivity Survey (EMSS) 
cluster arc survey by Le F\`evre et al. (1994)
is $8 \times 10^{44}\, h^{-2} {\rm ergs\, s^{-1}}$, measured in the
EMSS band  of 0.3 to 3.5 keV. Converting this to a total mass
by following through a recently derived $L-T$ relation (Arnaud \&
Evrard 1998) in addition to the above $M-T$ relation suggest a reliable lower
limit on the mass of $7.5 \times 10^{14}\, h^{-1}\, {\rm M_{\sun}}$.
We have taken the lowest limit on mass by considering all
cosmologies -- since the $M-T$ relation and $L$ estimates are
different under varying cosmologies.
The current observed arc statistics (e.g., Le F\'evre et al. 1994; Luppino
et al. 1998), when converted to a whole sky number, suggest
that the number of arcs towards above defined massive clusters
and with amplifications greater than 10 down to a V-band limiting
magnitude of 22 is between 1500 and 2500 (e.g., C99; Bartelmann et
al. 1998).  Ignoring the upper value, which is likely to be
unreliable, we use the lower estimate to derive an upper limit on $\Omega_m$. 
Since the lower estimate is based on the observed number,
this allows us to put a reliable upper limit on $\Omega_m$.
We also vary this lower limit to study its effects on our constraints.

In order to derive a constraint on $\Omega_m$, we
adopt a Bayesian approach, and take a uniform prior for
$\Omega_m$  between 0 and +1. This is primarily due to the fact that
we do not yet have a precise determination of
$\Omega_m$, and, based on various theoretical arguments, 
we do not wish to consider cosmologies in
which either this quantity lies outside the interval [0,1].
Since the prior for $\Omega_m$ is uniform,
the posterior probability density is simply proportional to the
likelihood.

The likelihood ${\cal L}$ --- a function of $\Omega_m$ --- 
is the probability of the data, given $\Omega_m$. 
The likelihood for $n$ observed arcs (at redshifts
$z_j$) when $\bar N$ is expected is given by (Cooray, Quashnock \&
Miller 1999):
\begin{equation}
\langle {\cal L}(n)\rangle =\prod_{j=0}^n \tau(z_j) \times e^{- \bar N} \times 
\left(1 + \sigma_{\tau}^2 \left[ \frac{\bar N^2}{2}-n\bar N +\frac{n(n-1)}{2}\right]\right) \; .
\end{equation}
We have taken into account the uncertainty in predicted lensing rate
by introducing $\sigma_{\tau}$, which is the fractional 1-$\sigma$ 
error on $\tau$ and then mariginalizing the likelihood
over the variance of it. In order to constrain $\Omega_m$, we also need the 
redshifts $z_j$ of the observed arcs. Using published data on
individual lensing clusters, we obtained a median redshift for the lensed
arcs of $\bar z \sim 1.6$.  As we find below, 
changing this mean redshift to a 
reasonably different value does not change our constraints on 
the $\Omega_m$ greatly. 

When the observed lower limit is 
compared with predictions, we find that $\Omega_m \lesssim 0.62$ at
the 95\% confidence when there is no core radius
(SIS). When we include a core radius of 35 h$^{-1}$ kpc, the upper
limit on $\Omega_m$ decreases to 0.56 at the 95\% confidence level; The change
in the upper limit on $\Omega_m$ is only a minor effect.
If the core radius were to be as large as 100 h$^{-1}$ kpc,
then the derived upper limit on $\Omega_m$ can be as low as 0.29 at the
95\% confidence level. 
However, such a large core radius
for the cluster dark matter profile is ruled out leaving the
possibility
only for a much smaller core radius of the order 30 to 40 h$^{-1}$ kpc.
When the effective median redshift of lensed
arcs are changed to a lower number as  $\sim$ 1, the upper limit
increases to 0.58 from 0.56, while when the redshift is increases to 
a value of 3.0 from 1.6, the upper limit on $\Omega_m$ at the 95\%
confidence decreases to 0.52.
When we increase the lower limit from 1500 to 2000, our upper limit on
$\Omega_m$ with a model involving core radius of 35 h$^{-1}$ kpc,
decreases to 0.52 from 0.56. This is primarily due to the fact that
the expected number of lensing events varies by orders of magnitude when
$\Omega_m$ is changed from 1 to 0, with the variation in the
expected number larger at the lower end of $\Omega_m$ values.
For such a small core radius, the limit on $\Omega_m$ is consistent with
current estimates based on other cosmological probes such as
type Ia supernovae and galaxy cluster abundances. 
We note that our limit on $\Omega_m$ does not mean that the Universe
is open without  a cosmological constant, but rather lensed arc statistics
are not sensitive enough to the cosmological constant
to see its effects above the current uncertainties.
In general, the upper limit on $\Omega_m$ with a cosmological constant
is slightly higher when compared to an open model. However, this
difference is rather small ($\sim$ few percent, see
Fig.~1), and cannot be distinguished using 
current observations on cluster number counts and lensed arcs.

\subsection{Uncertainties \& Systematic Effects}

Using a lower limit on the observed number of lensed arcs, we have
derived an upper limit on $\Omega_m$. A major uncertainty is likely to
come when  estimating a lower limit on the 
observed number of arcs since it is only based on
optical followup observations of EMSS clusters
(e.g., Henry et al. 1992). As a reliable
approach, we have taken the lower limit allowed by the observed number
of lensed arcs towards this sample. 
In reality, the true number is likely to be higher but the lower
limit allows us to safely consider upper limits on cosmological
parameters, especially the cosmological mass density.
The present observational number on the number of lensing events is unlikely to
be improved unless large samples of clusters are followed up
at optical wavelengths. Several attempts are currently underway (e.g.,
Luppino et al. 1998), however,
all such surveys our still  based on the EMSS sample. It is likely
that the optical followup observations of additional
cluster catalogs, such as the ROSAT Bright Cluster Survey (BCS; Ebeling et
al. 1998), can greatly improve our knowledge on the lensing statistics
due to galaxy clusters allowing better constraints on the cosmological
parameters. 

In addition to current low number statistics, 
other uncertainties are likely to come
from the conversion of observations, such as cluster X-ray luminosity,
to mass. However, at each step, we have considered various estimates
such that the predicted number of lensed arcs is overestimated;
This approach allows us to consider a reliable limit on $\Omega_m$,
whose upper limit may have been systematically increased by our procedure.
We have also investigated the effect of a core radius on 
arc statistics. As found, for luminous optical arcs with
amplifications greater than 10, the effect of a core radius on our
prediction on the  number of lensing events is minimal. The upper
limit only varies from 0.62 to 0.56 at the 95\% confidence when
a reasonable core radius of size 35 h$^{-1}$ kpc is introduced.
Increasing the core radius as high as 100 h$^{-1}$ kpc reduces
the upper limit by a factor of $\sim$ 2, however, such a large core
radius is ruled out by current observations of gravitational lensing
of clusters (e.g., the nonexistence of radial arcs at large distances
from the cluster center).

\section{Summary \& Conclusions}

Using a lower limit on the observed number of lensed arcs due to
clusters, we have calculated an upper limit on $\Omega_m$.
Due to large uncertainties in the predicted number of lensed sources,
primarily dominated by the error in $\sigma_8$, we are unable to
place limits on the cosmological constant. However, after considering
possible known errors, and carefully taking account various estimates
such that the upper limit on $\Omega_m$ is not reduced, we
conclude that $\Omega_m \lesssim 0.62$ at the 95\% confidence.

\acknowledgments
I am grateful to Richard Mushotzky for pointing out the possibility
to derive an upper limit on $\Omega_m$, Ian
Dell'Antonio for communicating details of his numerical inversions
and constraints on cluster density profiles.
I also acknowledge useful comments from an anonymous
referee which led to several improvements in the paper and acknowledge
partial support from a McCormick Fellowship at the University of Chicago.

\end{document}